\documentstyle[osa,manuscript,psfig]{revtex}

\begin{document}

\titlepage
\title{
Spin content of Lambda and its longitudinal 
polarization in $e^+e^-$ annihilation at high energies } 
\author {C. Boros$^{1,2}$, and Liang Zuo-tang$^{1,3}$}
\address {$^1$Institut f\"ur Theoretische Physik,
Freie Universit\"at Berlin,
Arnimallee 14, 14195 Berlin, Germany\\
$^2$Special Research Centre for the Subatomic Structure of Matter,
University of Adelaide, Adelaide, Australia 5005\\ 
$^3$Department of Physics,
Shandong University,Jinan, Shandong 250100,China}

\maketitle

\begin{abstract}                

Longitudinal polarization of Lambda produced in 
$e^+e^-$ annihilation at LEP energies  
is calculated in a picture for the spin content of Lambda  
which is consistent with the polarized deep inelastic 
lepton-nucleon scattering data and SU(3) flavor symmetry 
for hyperon decay so that 
the spin of Lambda is not completely 
carried by its $s$-valence quark. 
A comparison with the recent ALEPH data 
and the results of earlier calculations 
based on the static quark model in which 
the spin of Lambda is completely determined 
by the $s$-quark is given.   
The result shows that further measurements of such polarization 
should provide useful information to the question 
of which picture is more suitable in describing the spin effects in 
the fragmentation processes. 

\end{abstract}     

\newpage

There exist now in literature two completely different 
pictures for the spin contents of the baryons:
According to the static (or constituent) quark model, 
spin of a baryon belonging to the $J^P={1\over 2}^+$ 
octet is completely determined by 
the three valence quarks. 
The sum of the spins of these valence quarks 
is the spin of the baryon. 
This picture is very successful in describing 
the static properties of the baryons.
But according to the 
the data from polarized deep inelastic 
lepton-nucleon scattering [1]
and SU(3) flavor symmetry in hyperon decay, 
the sum of the spins of the 
three valence quarks is only 
a small fraction of the 
spin of the nucleon.  
A large part of the baryon spin originates from the 
orbital angular momenta of the valence quarks and/or 
from the sea (i.e. the sea quarks, antiquarks and gluons).
Hence, it is natural to ask
which picture is suitable in describing 
the spin effects in the quark fragmentation process.   
Obviously, the answer to this question is a priori unknown
and should be studied in experiments. 
Polarization of Lambda is an ideal place to 
investigate this problem because of the following:  
First,  the spin structure of Lambda 
in the static quark model is very special:  
the spin of Lambda is completely carried by 
the $s$ valence quark 
while the $u$ and $d$ have completely no contribution.
This picture is completely different from that 
drawn from the data of 
deep-inelastic lepton-nucleon scattering [1] 
and SU(3) flavor symmetry in hyperon decay.   
The deep inelastic scattering data, 
together with the SU(3) flavor symmetry for hyperon decay,
suggest that [2] the $s$ quark 
carries only about $60\%$ of the Lambda spin, 
while the $u$ or $d$ quark each 
carries about $-20\%$. 
Second, the polarization of the produced Lambda 
can easily be determined 
in experiments by measuring the angular distribution 
of the decay products. 
Besides, striking polarization effects have been observed 
for hyperons  produced in unpolarized 
hadron-hadron collisions experiments [3].
Such effects have been observed  
for more than two decades and remain as 
a puzzle for the theoretians. 
Clearly, the study of the above mentioned 
question should be able to provide some useful 
information of this problem; and this 
makes the study even more 
interesting and instructive. 

Polarization effects for Lambda 
produced in high energy reactions
have been studied in different connections [2,4-12].
In some of these discussions [2,4-9], 
current quark picture has been used 
thus the picture for the spin content of Lambda 
drawn from the polarized deep 
inelastic lepton nucleon scattering data 
should be applicable.
But in the other [10-12], 
it is assumed that Lambda spin is completely determined by the 
$s$ quark thus picture of the static quark model 
should be applicable. 
No discussion has been made yet to the question 
of which of them is more suitable.

It is known from the standard model of 
electroweak interaction that 
the $s$ quark produced in $e^+e^-$ annihilation 
at high energies is longitudinally polarized [13]. 
Hence it is expected [13] 
that the Lambda which contains this $s$ quark 
should also be longitudinally polarized 
and such Lambda polarization can be measured in experiments. 
Theoretically, this Lambda polarization can be calculated 
and the results should be quite different using 
the above mentioned two different pictures 
for the spin contents of Lambda. 
Hence, measurements of the polarization should be able to 
show which picture is more suitable in describing such spin effects.
A calculation of the longitudinal 
Lambda polarization in $e^+e^-$ 
annihilation at the $Z$-pole 
has been made [14] using the picture of 
the static quark model, but no calculation has been made 
yet [15] using the picture drawn from the data of 
deep inelastic scattering.

More recently, longitudinal Lambda polarization 
in $e^+e^-$ annihilation at the $Z$-pole 
(which is therefore dominated by those from $Z$ decay) 
has been measured [16] 
by the ALEPH Collaboration at CERN. 
A comparison of the data [16] with 
the calculated results of [14] has been made [16], 
and they are in good agreement with each other.
This means the above mentioned static quark model picture 
for Lambda spin structure is 
consistent with the data [16]. 
Does this mean that the static (or constituent) quark model 
but not that from deep inelastic lepton nucleon scattering 
should be used in the fragmentation process?  
To answer this question, 
calculations have to be carried out 
using a picture which is consistent with 
the deep inelastic scattering data 
so that a comparison with the ALEPH data [16] 
can be made.

In this note, we present the results of 
such a calculation 
and compare them with those obtained in [14] 
and the ALEPH data [16]. 
The calculations have been carried out 
using the same method as that in [14]. 
Here, we first consider the contribution of the Lambdas 
which are directly produced in the hadronization process.
Such hyperons are divided into two groups: 
those which contain the leading $u$, $d$ or $s$ quark 
and those which do not. 
The latter kind of Lambdas, i.e. those which do not contain 
the initial $u$, $d$ or $s$ quark from $e^+e^-$ annihilation, 
are assumed [14] not to be polarized [17] but the 
former kind can be polarized since 
the initial $u$, $d$ or $s$ quark is 
longitudinally polarized.
The polarization of such Lambda is different  
in different pictures for the spin structure of Lambda. 
More precisely, the polarization of such Lambda 
is equal to the fraction of spin 
carried by the quark which has the flavor of the 
initial quark multiplied by  
the polarization of this initial quark. 
The polarizations of the initial quarks from 
$e^+e^-$ annihilations are determined 
by the standard model for electroweak interactions, 
and given by [13],
\begin{equation} 
P_f=-\frac{A_f(1+\cos^2\theta)+B_f\cos\theta}
          {C_f(1+\cos^2\theta)+D_f\cos\theta}
\end{equation}
where $\theta$ is the angle between the outgoing quark and 
the incoming electron, the subscript $f$ denotes the flavor 
of the quark and
\begin{equation}
A_f=2a_fb_f(a^2+b^2)-2(1-{m_z^2 \over s})Q_fab_f,
\end{equation}
\begin{equation}
B_f=4ab(a^2_f+b^2_f)-2(1-{m_z^2 \over s})Q_fa_fb,
\end{equation}
\begin{equation}
C_f=\frac{(s-m_Z^2)^2+m_Z^2\Gamma^2_Z}{s^2}Q^2_f+
(a^2+b^2)(a^2_f+b^2_f)-2(1-{m_z^2 \over s})Q_faa_f,
\end{equation}
\begin{equation}
D_f=8aba_fb_f-4(1-{m_z^2 \over s})Q_fbb_f,
\end{equation}
where $m_Z$ and $\Gamma_Z$ are the mass and decay width of $Z$; 
$a, b, a_f$ and $b_f$ 
are the axial and vector coupling constants 
of electron and quark to $Z$ boson,  
which are functions of the Weinberg angle $\theta_W$. 
(See table 1 in [13]). 
Averaging over $\theta$, we obtain 
$P_f=-0.67$ for $f=u,c,t$ and $P_f=-0.94$  for $f=d,s,b$. 

The fractional contributions 
($\Delta U_\Lambda$, 
$\Delta D_\Lambda$, and 
$\Delta S_\Lambda$) 
of different flavors ($u$, $d$ and $s$) 
to Lambda spin are calculated using 
the deep inelastic lepton-nucleon scattering data 
on $\Gamma_1\equiv \int^1_0g_1(x) dx$ 
[where $g_1(x)$ is the spin-dependent structure]
and those for the constants $F$ and $D$ in hyperon decay. 
The detailed procedure of extracting the 
$\Delta U_\Lambda$, 
$\Delta D_\Lambda$, and 
$\Delta S_\Lambda$ from the data for $\Gamma_1^p$ for proton, 
and those for $F$ and $D$ is summarized in the Appendix.
The obtained results are given in Table 1.

We next consider the contribution 
of those Lambda's from 
the decay of other hyperons in the same octet as Lambda. 
These hyperons can also be polarized if they contain the 
initial $u$, $d$ or $s$ quark, and the polarization 
can be transferred to Lambda's in the decay processes.
The polarization of such Lambda is thus equal to the 
polarization of the hyperon multiplied by the probability for 
the polarization to be transferred to Lambda. 
Hence, to calculate such contribution, 
we need to calculate the polarization 
of the such hyperon before it decays and 
the probability for the polarization 
to be transferred to Lambda in the decay process. 
The polarization of hyperon in the same octet as Lambda 
can easily be calculated using exactly the same method 
as that for Lambda. 
There are three such hyperons, i.e. 
$\Sigma^0$, $\Xi^0$ and $\Xi^-$ which may decay into $\Lambda$.
We calculated the fractional contributions 
of different flavors of quarks to their spins 
in the way described in the Appendix and obtained 
the results shown in Table 1.
These results are as reliable as those for Lambda, 
and are therefore [2] as reliable as those for the nucleons.
$\Sigma^0$ decay into $\Lambda$ by emitting a photon, i.e.,  
$\Sigma^0\to \Lambda \gamma$. 
Whether the polarization of $\Sigma^0$ is 
transferred to the produced Lambda in this 
decay process has been discussed in [18]. 
It has been shown that, on the average, 
the produced $\Lambda$ is also polarized 
(in the opposite direction as $\Sigma^0$)
if $\Sigma^0$ was polarized before its decay,  
and the polarization is $-1/3$ of that of the $\Sigma^0$. 
The hyperon $\Xi$ decays into $\Lambda$ through 
$\Xi \to \Lambda \pi$, which is a parity 
non-conserving decay and is dominated by S-wave. 
The polarization of the produced $\Lambda$ is equal to 
that of the $\Xi$ multiplied by a factor $(1+2\gamma)/3$, 
where $\gamma$ can be found in review of particle properties [19] 
as $\gamma = 0.87$.

It is now still impossible to calculate 
the polarizations of the produced 
hyperons that belong to the baryon decuplet 
in a way consistent with that for those in the octet. 
This is because no deep-inelastic 
scattering data on any one of such baryons is available.
It is therefore impossible to calculate 
the fractional contributions 
of different flavors to the spin of such hyperon. 
Hence, it is impossible to estimate the
contributions of decays of such hyperons 
which contain the initial $u$, $d$ or $s$ quark 
to the polarization of Lambda in 
the final state of $e^+e^-$ annihilations 
in the same way as that for the octet hyperons. 
Qualitative analysis suggests that
the influences of such hyperons 
should not be very large.
This is because, first, their production rates 
are relatively small, and  
second, since the mass differences 
between such hyperons and Lambda are relative large, 
their decays contribute mainly to 
Lambda's in the central region of the $e^+e^-$ annihilation 
(i.e. those with relatively small momenta).  
This region is dominated by those Lambda's 
which do not contain the initial quark 
and are unpolarized.

To make a quantitative estimation, we need a hadronization model 
to calculate all the different contributions to the Lambda's 
from all the different sources discussed above. 
For this purpose, we used, as in [14], 
the LUND model [20] as implemented by JETSET [21]. 
We explicitly calculated the different contributions, 
and obtained the results shown in Fig.1. 
We see in particular that the contribution 
from the decay of the decuplet hyperons 
is indeed relatively small. 
We calculated Lambda polarization $P_\Lambda$ 
for the following two cases: 
In the first case, we completely neglect  
the contribution from decuplet hyperon decay to $P_\Lambda$  
and obtained the results shown by the 
solid line in Fig.2. 
In the second case, we used the results 
for the polarization of the decuplet hyperons 
obtained from the static quark model 
as an approximation to estimate the contribution 
of such hyperon decay to $P_\Lambda$. 
We added the results to $P_\Lambda$ and obtained 
the dashed line in Fig.2. 
For comparison, we included  in the figure 
also the results from the static quark model 
without (dotted line) or with (dash-dotted line) 
the contributions from decuplet hyperon decay.

>From these results, we see that 
there is indeed a significant difference between 
those obtained in [14] based on 
the picture of the static quark model
and those obtained in the present estimation using a picture 
based on the polarized deep-inelastic 
lepton-nucleon scattering data [1] 
and SU(3) flavor symmetry for hyperon decay.
It seems that the ALEPH data [16] 
favors the former but cannot exclude the latter 
since the error bars are still too large. 
We see also that, although the influence from the decuplet is 
indeed relative small, but it is not negligible 
in particular for moderate $z$. 
We can also see that further measurements of $P_\Lambda$ 
with higher accuracy are needed to distinguish 
between these two kinds of models. 
The large $z$ region is most suitable for such a study  
since in this region not only the magnitude 
of $P_\Lambda$ itself is 
large but also the difference 
between the prediction of the two different models is large. 
It will be also particularly helpful to measure the 
polarization only for those Lambda's 
which are not decay products of decuplet hyperons. 

We thank Meng Ta-chung and R. Rittel for discussions. 
This work was supported in part 
by Deutsche Forschungsgemeinschaft 
(DFG Me7-1), FNK of FU Berlin (FPS Cluster),  
the National Natural Science Foundation of China 
and the Australian Research Council.

\vskip 1.0cm
\noindent
{\large Appendix} 

\vskip 0.2cm

The way of extracting the fractional contributions 
of quarks of different flavors 
to the spin of a baryon in the 
$J^P={1\over 2}^+$ octet 
from $\Gamma_1^p\equiv \int^1_0 g_1^p(x) dx$ 
obtained in deep-inelastic 
lepton-proton scattering experiments 
and the constants $F$ and $D$ obtained 
from hyperon decay experiments 
is now in fact quite standard [1]. 
We present in this appendix 
the main ingredients used
in this procedure in order to 
remind the readers of the assumptions one 
uses here and to show how one extends 
it to obtain the results for other baryons 
in the same octet as Lambda.

According to the quark parton model [22], we have 
\begin{equation}
  g_1(x)=\frac{1}{2}\sum_q e_q^2 [\Delta q(x) +\Delta \bar q(x) ],
\end{equation}
where $\Delta q(x)=q^{+}(x) - q{-}(x) $, 
$\Delta \bar q(x)=\bar q^{+}(x) - \bar q^{-}(x) $
is the difference between the number density of quarks 
(antiquarks) of flavor $q$ polarized 
in the same, and 
that of those polarized in the opposite, 
longitudinal direction as the nucleon;
$e_q$ is the electric charge of the 
quark in unit of electron charge.
Denoting $\Delta Q \equiv 
\int^1_0 [\Delta q(x) + \Delta \bar q(x)] dx$,  
we obtain, 
\begin{equation}
\Gamma_1=  \int_0^1 g_1(x) dx = 2\sqrt{\frac{2}{3}} \Delta Q_0 
                      + \frac{1}{6} \Delta Q_3 
                      +\frac{1}{6\sqrt{3}}\Delta Q_8, 
\end{equation}  
where $\Delta Q_0\equiv \frac{1}{12}\sqrt{2\over 3}
     (\Delta U +\Delta D +\Delta S)$; 
$\Delta Q_3\equiv \frac{1}{2} (\Delta U -\Delta D)$, and 
$\Delta Q_8\equiv  \frac{\sqrt{3}}{6} (\Delta U + \Delta D -2 \Delta S)$  
are the singlet, triplet, and octet terms. 
The singlet term $\Delta Q_0$ is proportional to the fraction 
$\Sigma=\Delta U+\Delta D+\Delta S$ 
of spin of the nucleon carried by the light quarks.  
If SU(3) flavor symmetry is hold, 
$\Sigma$ should be the same for 
baryons in the same SU(3) multiplet.
Using the method of operator product expansion, 
one relates the $\Delta Q_a$'s 
to the  matrix elements of local operators 
$A^\mu_a=\bar q \gamma_\mu \gamma_5 
 \frac{\lambda^a}{2} q $ by 
\begin{equation}
 2 M S^\mu \Delta Q_a = \langle P,S| A^\mu_a |P,S\rangle
\end{equation}
where $a=0,3,$ or $8$; 
$S$, $P$ and $M$ are spin, 
momentum and mass of the baryon $B$
and $\lambda^a$ are the usual SU(3) matrices 
acting in the flavor space.   
The matrix elements of the local operators 
are at $Q^2=0$ and can be measured in hyperon decays. 
Under the assumption that SU(3) flavor symmetry 
is valid among the baryons,  
one can use the Wigner-Eckart theorem and obtains [23],
\begin{equation}
\langle \psi_b | A^\mu_a|\psi_c\rangle = 
 2MS^\mu (i f_{abc} F + d_{abc} D)
\end{equation}
where $\psi_b$ and $\psi_c$ are the basis of the 
eight dimensional representation of SU(3) with spin $S$;
$f_{abc}$ and $d_{abc}$ are the totally antisymmetric and 
symmetric structure constants for SU(3) group; 
and the quantities $F$ and $D$ are constants which are
independent of the particular states 
in the same multiplet. 
 
The wave functions of the baryons $B$'s 
in the baryon octet can be expressed in terms of the  
basis vectors $\psi_a$ of the eight 
dimensional representation of SU(3). For example, 
$ |p\rangle = \frac{1}{\sqrt{2}} (\psi_4-i\psi_5) $,
$ |\Lambda\rangle = \psi_8,$
$ |\Sigma^0 \rangle = \psi_3,$
$ |\Xi^0\rangle = \frac{1}{\sqrt{2}} (\psi_6+i\psi_7),$
$ |\Xi^-\rangle = -\frac{1}{\sqrt{2}} (\psi_4+i\psi_5)$.
Using these wave-functions and Eq.(9) we can  
calculate the matrix elements of the 
axial currents between the different baryons in the octet and   
obtain the $\Delta Q_3$ and $\Delta Q_8$  as functions of $F$ and $D$. 
In this way, we obtain $\Delta Q_3^p= \frac{1}{2}(D+F)$ and 
$\Delta Q_8^p=\frac{1}{2\sqrt{3}}(3F-D)$. 
Using the experimental data for $F$ and $D$, 
and that for $\Gamma_1^p$, we obtain 
from Eq.(11) the $\Delta Q_0$ thus $\Sigma$ for proton,
which should be the same for all the baryons in the octet. 
We then use this $\Sigma$ and data for $F$ and $D$ 
to calculate the $\Delta U$, $\Delta D$ and $\Delta S$ for 
other baryons. 
The expressions of $\Delta U$, $\Delta D$ and $\Delta S$ 
in terms of $\Sigma$, $F$ and $D$, and their numerical results 
obtained using the data are listed in Table 1.

\widetext

\begin{table}
\caption{Fractional contributions $\Delta U$, $\Delta D$ and $\Delta S$ 
of the light flavors to the spin of baryons in the $J^P={1\over 2}^+$ 
octet calculated  
using the static quark model (static QM) and those obtained using 
the data for deep inelastic lepton-nucleon scattering and those for 
hyperon decay under the assumption that SU(3) flavor symmetry is valid.  
The first column shows the obtained expressions in terms of 
$\Sigma$, $F$ and $D$. 
The static QM results are obtained by inserting 
$\Sigma=1, F=2/3$ and $D=1$ into these expressions 
and those in the third column are obtained by inserting 
$\Sigma =0.28$, obtained from deep inelastic lepton-nucleon 
scattering experiments [1], 
and $F+D=g_A/g_V=1.2573, F/D=0.575$  
obtained [19,24] from the hyperon decay experiments. }
\begin{tabular}{l||c|c|c||c|c|c}
\hline 
 &\multicolumn{3}{c||}{$\Lambda$} &\multicolumn{3}{c}{$\Sigma^0$}\\ \hline 
& & static QM & DIS data & & static  QM & DIS data \\ \hline
$\Delta U$ & $\frac{1}{3} (\Sigma-D)$  & 0    & -0.17 & 
             $\frac{1}{3} (\Sigma+D)$  & 2/3  & 0.36  \\ \hline  
$\Delta D$ & $\frac{1}{3} (\Sigma-D)$  & 0    & -0.17 & 
             $\frac{1}{3} (\Sigma+D)$  & 2/3  & 0.36  \\ \hline
$\Delta S$ & $\frac{1}{3} (\Sigma+2D)$ & 1    & 0.62  & 
             $\frac{1}{3} (\Sigma-2D)$ & -1/3 & -0.44 \\ \hline \hline 
 &\multicolumn{3}{c||}{$\Xi^0$} &\multicolumn{3}{c}{$\Xi^-$}\\ \hline 
&  & static QM & DIS data & & static  QM & DIS data \\ \hline 
$\Delta U$ & $\frac{1}{3} (\Sigma-2D)$  & -1/3 & -0.44 & 
             $\frac{1}{3} (\Sigma+D)-F$ & 0    & -0.10 \\ \hline  
$\Delta D$ & $\frac{1}{3} (\Sigma+D)-F$ & 0    & -0.10 & 
             $\frac{1}{3} (\Sigma-2D)$  & -1/3 & -0.44 \\ \hline
$\Delta S$ & $\frac{1}{3} (\Sigma+D)+F$ & 4/3  &  0.82 & 
             $\frac{1}{3} (\Sigma+D)+F$ & 4/3  &  0.82 \\ \hline 
\end{tabular}
\end{table}

\newpage 

\noindent
{\large Figures}
\vskip 0.6truecm
\noindent
Fig.1: Fractional contributions to Lambdas 
produced in $e^+e^-$ annihilation 
at LEP energy from different sources:  
The solid line denotes those Lambdas which are 
produced directly and contain the initial $u$, $d$ or $s$ quark; 
the dash-dotted and dashed lines are those from decay of octet 
($\Sigma^0$, $\Xi$) and decuplet hyperons ($\Sigma^*$, $\Xi^*$) 
which contain the initial quarks. 
$z\equiv 2p/\sqrt{s}$, where $p$ is the momentum of 
the produced Lambda and $\sqrt{s}$ is the total center of 
mass energy of the $e^+e^-$ system.

\vskip 0.3cm
\noindent
Fig.2: Longitudinal polarization of Lambda, $P_\Lambda$, 
from $e^+e^-$ annihilation at LEP energy as a function of $z$.
(See text for more details).


\begin {thebibliography}{99}
\bibitem{[17]} This is not only true in the popular hadronization 
              models such as LUND model [18] but also 
              consistent with the experimental observations 
              that both hyperon polarization in unpolarized 
              hadron-hadron collisions and left-right asymmetries 
              in inclusive production processes in 
              single spin hadron-hadron collisions in the 
              central rapidity region are consistent with zero 
              although they are quite large in the fragmentation 
              region. (See e.g. [3] and the references given there).
\end{thebibliography}
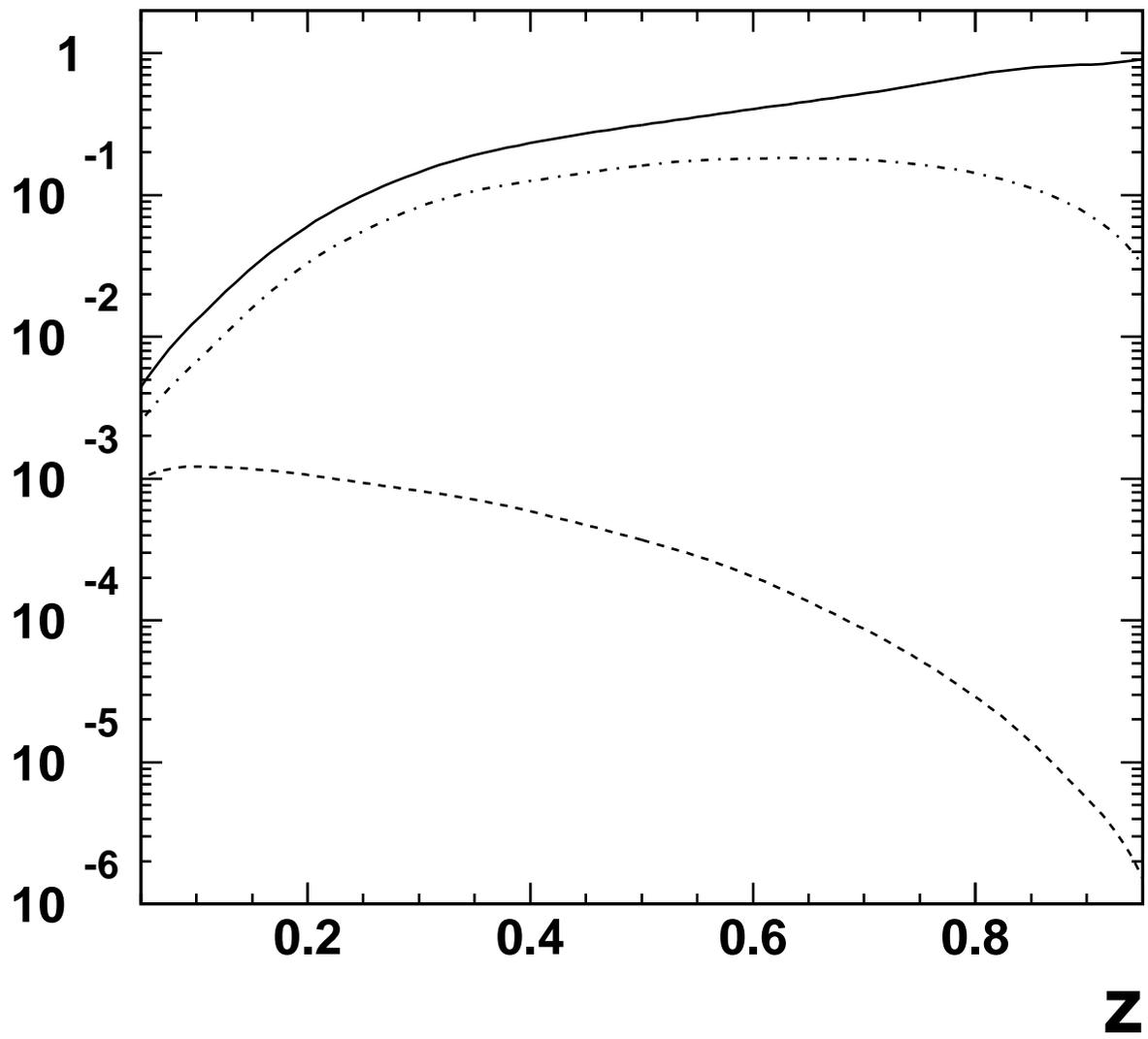
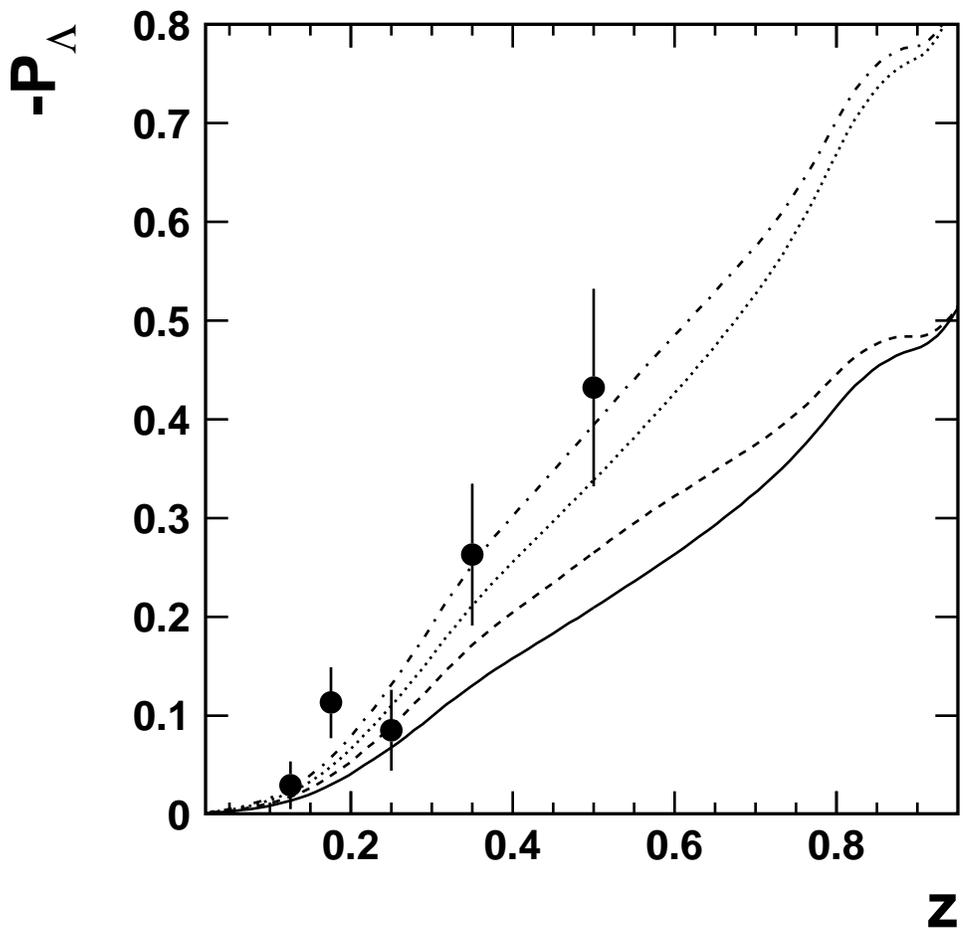
\end{document}